# Tuning color centers at a twisted interface

Cong Su[1,2,3,†], Fang Zhang[1,2,4,5,†], Salman Kahn[1,2,†], Brian Shevitski[1,2,3,6], Jingwei Jiang[1,2], Chunhui Dai[1,2,3], Alex Ungar[1,2,3], Ji-Hoon Park[7], Kenji Watanabe[8], Takashi Taniguchi[9], Jing Kong[7], Zikang Tang[5], Wenqing Zhang[4], Feng Wang[1,2,3], Michael Crommie[1,2,3], Steven G. Louie[1,2,\*], Shaul Aloni[6,\*], Alex Zettl[1,2,3,\*]

[1] Department of Physics, University of California, Berkeley, CA 94720, USA.

[2] Materials Sciences Division, Lawrence Berkeley National Laboratory, Berkeley, CA 94720, USA.

[3] Kavli Energy NanoSciences Institute at the University of California, Berkeley, CA 94720, USA.

[4] Department of Physics, Southern University of Science and Technology, Shenzhen 518000, China.

[5] Institute of Applied Physics and Materials Engineering, University of Macau, Macau S.A.R. 999078, China.

[6] The Molecular Foundry, Lawrence Berkeley National Laboratory, Berkeley, California 94720, USA.

[7] Electrical Engineering and Computer Sciences, Massachusetts Institute of Technology, Cambridge, MA 02139, USA.

[8] Research Center for Functional Materials, National Institute for Materials Science, 1-1 Namiki, Tsukuba 305-0044, Japan.

[9] International Center for Materials Nanoarchitectonics, National Institute for Materials Science, 1-1 Namiki, Tsukuba 305-0044, Japan.

[†] These authors contributed equally to this work.

[\*] Corresponding authors. Email: azettl@berkeley.edu (A.Z.), saloni@lbl.gov (S.A.), sglouie@berkeley.edu (S.G.L).

**Abstract**

Color center is a promising platform for quantum technologies, but their application is hindered by the typically random defect distribution and complex mesoscopic environment. Employing cathodoluminescence, we demonstrate that an ultraviolet-emitting single photon emitter can be readily activated and controlled on-demand at the twisted interface of two hexagonal boron nitride flakes. The brightness of the color center can be enhanced by two orders of magnitude by altering the twist angle. Additionally, a brightness modulation of nearly 100% of this color center is achieved by an external voltage. Our *ab-initio GW* calculations suggest that the emission is correlated to nitrogen vacancies and that a twist-induced moiré potential facilitates electron-hole recombination. This mechanism is further exploited to draw nanoscale color center patterns using electron beams.

Wide-band-gap materials such as diamond, silicon carbide, gallium nitride, and zinc oxide, two-dimensional (2D) semiconductors such as hexagonal boron nitride (hBN) and transition metal dichalcogenides, and one-dimensional nanotubes, have all been suggested as candidates for hosting color centers (*1*). These color centers may provide a basis for quantum computation, quantum information networks, and quantum sensors (*2*). Color centers in hBN are particularly noteworthy in that it displays emission that is bright (*3*, *4*) and stable at high temperature (*5*), and has optically addressable spin properties (*6*, *7*), with emission spectra spanning across ultraviolet (UV) (*8–11*), visible (*12–14*), and infrared spectral ranges (*6*). The van der Waals layered crystal structure of hBN makes it relatively easy to control the thickness through exfoliation (*15*), which is vital for color centers as the depth is critical (*16*). However, up to date, color centers are still generally randomly distributed, for example after fabrication by ion bombardment, and their on/off control is difficult (*17*, *18*). Though optical gating with a laser is an effective method to activate or deactivate color centers, as has been demonstrated on nitrogen vacancy centers in diamond and charged defects in hBN (*19*, *20*), neither full modulation of the emission amplitude, nor the brightness control by an external electric field, has been achieved. Having such control is essential for utilizing color centers in sensing and device applications.

Recently, moiré superlattices have led to new physical properties in twisted bilayer graphene, such as a Mott insulator, unconventional superconductivity (*21*, *22*), and ferromagnetism (*23*). In bilayer or multilayer semiconductor systems, moiré potentials can modify band-edge emission, creating new peaks in the photoluminescence spectrum (*24–27*). Electric dipoles (*28*, *29*) and non-linear optical responses (*30*) have been found at the twisted interface of hBN due to symmetry breaking. From an engineering perspective, the rotation between hBN flakes is reasonably accessible as a control parameter, as it can be continuously adjusted, for example by the cantilever of an atomic force microscope (*31*) or by a polymer handle (*32*).

In this work, we show that a UV-emitting color center can be collectively activated near twisted hBN interface. Using cathodoluminescence (CL) for spectrum acquisition and high-resolution transmission electron microscope (HRTEM) for twist angle assignment, we find the brightness of this color center can be enhanced by two orders of magnitude upon twisting compared with the zero-twisted case. We perform calculations employing the *ab initio GW* (*33*) and *GW* plus Bethe-Salpeter equation (*GW*-BSE) (*34*) methods to investigate the origin of this color center and to provide a physical understanding of the brightness enhancement. Our theoretical analysis leads to an experimental demonstration that an external applied voltage can completely turn off the color center and decrease the brightness of the entire detectable UV range (280–409 nm) by 20-fold. We also explain the mechanism of the twist-enhanced color center emission and prove it by creating color center ensembles with controllable intensity and spatial positioning using electron-beam irradiation. This work thus provides effective methods of activating color centers around a 2D plane, and, in addition to twist angle, further extends modulation capability to external voltage application and electron irradiation.

To study the effect of twist angle on light emission intensity, we prepare twisted hBN (T-hBN) samples: one with nearly 0° twist (Fig. 1A), and another with various twist angles (Fig. 1D). Figures 1B and 1E show the intensity of the CL signal integrated between 280 and 409 nm (referred to as "UV intensity map" hereinafter) collected by photomultiplier tubes (PMT). In Fig. 1B, the nearly 0°-twist stack shows no significant difference in UV light intensity between the stacked area and the pristine top and bottom hBN layers, whereas Fig. 1E displays increased brightness where the hBN top layers are twisted at 14.5°, 24.8°, and 28.0° with respect to the bottom layer. It is evident that the brightness is significantly higher in the two-layer hBN stacks and increases with the twist angle showing a 50-fold enhancement at 28°. Figures 1C and 1F schematically illustrate the observation of brightness enhancement due to lattice twist. We have also proved that the emission is localized around the twisted interface, which is shown in figs. S2 and S3 of Supplementary Materials (SM) (*32, 33*).

Analysis of the spectral signature of the emitted light indicates that it originates from a well-defined color center coupled to lattice phonons. Figure 2A compares CL spectra of the color center from single crystal and T-hBN areas acquired at 35 K. A typical hBN is dark with no emission detectable (yellow curve). The bright UV color centers, which are very sparsely distributed in bulk hBN, include the emission from the defect itself at 305 nm (4.06 eV), commonly referred to as a zero-phonon line (ZPL), and its phonon sidebands (PSBs) at 321, 336, 349 nm (3.86, 3.69, and 3.55 eV, respectively) in hBN (red curve). The PSBs are phonon replicas of ZPL by longitudinal optical phonon at the Γ point (*11*). The ZPL and PSB peak assignments are confirmed by temperature-dependent CL shown in fig. S4, where the intensity of ZPL increases and PSB decreases when temperature is lowered. Previous studies have demonstrated that this emission originates from a paramagnetic defect in hBN (*35*), which acts as a single photon emitter with a second-order correlation coefficient $g^{(2)}(0) = 0.2$ (*10*). Intensity of this emission is significantly enhanced in T-hBN (blue curve).

Figure 2B shows spectra for different twist angles from 0.02° to 25.5°. When the twist angle is 0.02°, the CL emission is very weak, with no significant peak found in the UV range. As the twist angle increases, the intensity of peaks [plotted using conventional non-negative matrix factorization (NMF), see Method in SM for details] between 300-400 nm becomes more prominent, while the energy of the 300-nm ZPL does not shift. Fig. 2C shows the brightness of the UV emission (280-409 nm) captured via PMT as a function of twist angle. Samples on both TEM grids (left axis) and SiO$_2$/Si wafers (right axis) show a similar trend, where the brightness at the interface has a minimum at 0° and is maximized between 20°–30°. In particular, the emission increases by a factor of 60.5 when the twist angle increases from 0.02° to 27.9°. It should be noted that the twist angle, $\theta$, determined by diffraction pattern in TEM corresponds to two possible configurations: the two layers can be rotated either from an AA' or AB stacking orders, which are indiscernible by diffraction pattern in a multilayer hBN.

Figures 2D and 2E show HRTEM images for 0.02° and 12.2° twisted samples respectively, as representatives of small and large twist angles. For small-angle twisting of 2D materials, crystal distortion and reconstruction is commonly observed (*36*). As a result, for the 0.02° T-hBN, we see honeycomb structure in the low-magnification image and a complete restoration of crystal stacking in the area other than the boundaries of hexagons (see the zoom-in atomic image). In the case of a large-angle twist, like the 12.2° shown here, crystal distortion is not observable, creating homogeneous moiré patterns across the entire T-hBN.

Several conclusions can be made at this point: (1) The 300 nm (or 4.1 eV) ZPL is insensitive to the twist angle and is significantly lower than the band-to-band emission in hBN typically observed at around 210 nm (or 5.9 eV) (*37*), suggesting it is not a band-edge emission. Our calculation also suggests that moiré potential cannot decrease the emission from 5.9 eV to 4.1 eV (see Fig. S8). (2) The emission intensity, on the other hand, strongly depends on the crystal structure (twist angle), suggesting the emission cannot be from the transition between two deep defect levels inside the band gap of hBN. (3) Compared with other spectrum ranges (inset in Fig. 2B), the intensity enhancement of the defect emission led by 300-nm ZPL is clearly the most prominent in our sample. (4) Based on the preponderance of emitters around the interface as observed in Fig. 1E, the defect related to the 300-nm ZPL appears to be a commonly existing defect type in hBN.

The origin of the 300 nm (or 4.1 eV) peak has been controversial for over a decade (*9, 10, 38–40*). Experimental results shed some light here. First, the emission intensity increases with a decrease of nitrogen source during synthesis (*39*). Second, the 300-nm ZPL appears with the presence of carbon in the precursor during synthesis (*35*). These two clues indicate that the nitrogen vacancies ($V_N$) and substitutional doping of carbon atoms to nitrogen atoms ($C_N$) or to boron atoms ($C_B$) are the most probable origins of the deep-UV emission at 4.1 eV (300 nm).

To gain deeper insight into the origin of the defect-driven emission, we adopt the *ab initio GW* (*33*) and *GW*-BSE (*34*) approach to calculate the quasiparticle (QP) energies and optical response of the defects, respectively, including electron-electron interactions and excitonic contributions. This approach has been applied with success to a wide variety of materials, including 2D systems with defects (*41*). Here, we consider three types of defect structures: $V_N$, $C_N$, and $C_B$, based on the above analysis. Fig. 2B indicates experimentally that the 300-nm ZPL barely shifts in energy with respect to twist angles, so it is safe to use the defect spectra from a *non-twisted* hBN calculation to discover the origin of the 4.1-eV emission in T-hBN. To model the isolated defect states, we use the supercell approach that contains approximately 200 atoms (see Method section in SM for detailed computational methods).

We present our calculated QP defect energy levels of charge-neutral systems within the bandgap in Fig. 3A. For each type of defect, there are two QP defect energy levels in the band gap, corresponding to spin-up (blue) and spin-down (magenta) channels. The spin-down QP defect level is occupied while the spin-up level is unoccupied, giving a local magnetic moment of 1 $\mu_B$ for all three defects considered. The calculated optical *absorption* spectra using the *ab initio GW*-BSE approach (which includes electron-hole interaction or excitonic effects) of bulk hBN containing $V_N$, $C_N$, or $C_B$, as well as that of pristine bulk hBN are shown in Fig. 3B. We validate our method by comparing the experimental band-edge absorption peak of bulk hBN from CL (5.81 eV) (*37*) with our theoretical result (5.79 eV). The match between theory and experiment is extremely good, illustrating the necessity of including many-electron interaction effects in the *ab initio* calculations. For each defect configuration, the lowest-energy optical bright exciton is denoted by arrows. These excitons should give rise to the peaks that will be the most prominent in *emission* spectra after excitation of the system (*42*). There are two kinds of optically active excitons of distinct character for each type of defect, corresponding to a correlated electron-hole state with: (i) hole in the spin-down QP defect levels and an electron in the hBN spin-down intrinsic conduction states, and (ii) a hole in the hBN spin-up intrinsic valence states and an electron in the spin-up QP defect levels. Interaction between these two kinds of excitons is expected to be very weak owing to the spin consideration; thus, the lowest-energy bright exciton of each kind would be seen in an emission spectrum. We find that the lowest-energy bright exciton formed between the spin-up defect state in $V_N$ and the hBN intrinsic spin-up valence states has an energy of 4.3 eV with significant oscillator strength (also indicated by arrow in Fig. 3B), in good agreement with the 4.1-eV or 300-nm ZPL observed. In contrast, the other commonly proposed defects, $C_N$ and $C_B$, fail to yield any lowest-energy optically active excitons, of either kind (i) or kind (ii), with energy in the range from 2.5 to 4.9 eV. Therefore, we may rule them out and consider $V_N$ as the origin of the 4.1-eV or 300-nm ZPL.

Two points should also be noted: (1) Even though previous experiments have shown that the 4.1-eV peak is correlated with the concentration of C atoms in the precursor (*39*, *43*), we believe that a C dopant (especially $C_N$ in a N-deficient environment) is not directly involved in the emission process, but that it only encourages more $V_N$ to be created. In the Kröger-Vink notation, $V_N^{\cdots}$ has triple positive charges and $C_N'$ has a single negative charge. As a result, more $C_N$ simply leads to more $V_N$ due to the balancing of charges to make the whole system charge-neutral. (2) In our *ab initio GW* calculations, we assume a fixed lattice for the ground and excited electronic states, which essentially ignores the electron-phonon coupling. The rigid lattice approximation is good for screening defect candidates; however, we expect the predicted emission energy to be even closer to the experimental value if lattice relaxation is considered, as it would essentially lower the energy of ZPL.

Since according to our calculations, the 4.1-eV ZPL is from exciton states formed with an electron in the $V_N$ defect states and a hole in the hBN intrinsic valence states, the defect emission should have a strong correlation with the excited carrier concentration of the hBN crystal. We confirm this by showing that the brightness of the 4.1-eV emission can be modulated by extracting the excited charge carriers from the system via application of an electric field. Fig. 3C shows the vertical layout of the device. The T-hBN sample is encapsulated by two multilayer graphene sheets which allows for the application of a perpendicular electric field $E_t$ across the sample, via application of an external voltage $V_t$. A backgate voltage between the substrate and the bottom electrode can be tuned independently by $V_b$. $E_t$ can drive the electron-beam excited mobile carriers in T-hBN to the two electrodes, giving rise to electron-beam-induced current (EBIC). The EBIC, collected by the electrodes, decreases the carriers available for radiative recombination, thus decreasing defect emission. Indeed, we show that the 300 nm emission is strongly correlated to $V_t$: When $V_t$ is increased, either positively or negatively, the defect emission is greatly suppressed, as expected (Fig. 3D). The 300 nm emission becomes completely unobservable after $V_t$ exceeds 20 V (Fig. 3E). Only two minor peaks located at 330 nm and 400 nm remain emitting at high $V_t$, suggesting that these peaks most likely originate from recombination of excitons associated to deep levels, rather than involving excited carriers from the band edges. This statement is also supported by time-gated luminescence studies done previously (*44*). In Fig. 3F, the anti-correlative relationship of the UV intensity and EBIC as functions of $V_t$ shows that the emission of 300-nm ZPL with its PSB shares a common source with EBIC, which all involve the participation of excited carriers in hBN, making them complementary to each other (Fig. 3G). When the amplitude of $V_t$ is above 20 V, a nearly 100% modulation of intensity is achieved for the 300-nm ZPL. In the entire detectable UV range (280–409 nm), about 95% modulation can be reached, with the emission from the two deep-level-state excitons accounting for the rest 5%. On the other hand, the emission is completely insensitive to the backgate, as expected for a localized emission at the twisted interface (fig. S7).

The enhancement of emission coming from the twist interface originates from what we denote as "trap-state saturation", as a result of electronic band structure changes due to moiré potential. In our theoretical analysis, we find that twisting a bilayer hBN system lowers the conduction band minimum (CBM) by 0.2 eV at 21.78°. The valence band maximum (VBM), in contrast, is independent of lattice twisting (see Fig. 4A and fig. S9 for more twist angles). The constant energy difference between the QP defect state of $V_N$ and VBM explains the unchanged position of 4.1-eV ZPL at different twist angles in Fig. 2B. In addition, as previous temperature-dependent luminescence experiments have revealed, there are multiple trapping defect states (from different origins) located from 0.1 eV to 0.3 eV below CBM in bulk hBN which directly limit the emission of 4.1-eV ZPL (*35*, *39*). In larger-twisted-angle T-hBN, a substantial proportion of those trapping states would be energetically above the CBM in the interface region, losing their ability to trap electron carriers located at the CBM locally (Fig. 4B). As a result, the electron-beam excited electrons in the CBM could be more efficiently scattered into the unoccupied $V_N$ defect level, forming excitons with the holes in VBM which leads to stronger emission intensity. It should be noted that the detailed physical origin of these shallow trapping states remains to be explored, but we may tentatively propose one shallow state of $V_NC_B$ as a candidate (fig. S10).

An interesting corollary of the above mechanism follows: if we can intentionally saturate the trap states in hBN by a constant external stimulus, say an electron beam, the 4.1-eV emission should also be enhanced as a result. Indeed, we show that this is the case. In Fig. 4C, the color center ensembles with 300-nm ZPL are created by parking an electron beam for various dwell times (spectrum shown in fig. S11), and the brightness of UV emission is linearly proportional to the electron dose, though different saturation rates are observed under and above 2.5 s (fig. S12). Controlled creation of the color center ensembles can be achieved with a high spatial precision using a guided electron beam, as shown in Fig. 4D, where we draw a smiley face on a single crystal hBN consisting of 37 ensembles with the same intensity (except for the one on the right, which has a longer exposure time). The mechanism of trap state saturation using electron beams is shown in Fig. 4E, where we compare the cases of unsaturated trap states before electron beam exposure (left) and saturated trap states after a substantial electron beam exposure (right). We find that the color center ensembles activated by an electron beam can be annihilated by thermal annealing at 850 °C, where the trapped electrons can be released from the trap states thermodynamically, in agreement with thermoluminescence measurements (*35*). As a comparison, the emission at the twisted interface is stable under thermal annealing and under electron beam doping, a direct consequence of the intrinsic saturation (or inactivation) of trap states in T-hBN.

In conclusion, we demonstrate that a color center with 300 nm (or 4.1 eV) emission in hBN can be controlled by three different methods: (1) twisting the hBN layers enhances the emission localized around the twisted interface by orders of magnitude, (2) applying an external voltage can modulate the amplitude of emission in the UV range (280–409 nm) by 95%, and the 4.1-eV ZPL by nearly 100%, and (3) electron beam irradiation in hBN can continuously enhance the 4.1-eV color center ensembles by saturating trap states. The defect responsible for this emission is proposed to be that of $V_N$, which arises from a "defect level-valence band" exciton according to *ab initio GW*-BSE calculations. We explain the peak intensity enhancement in large-angle T-hBN as given by an attractive effective potential on the conduction band states created by the interface, where the formation of excitons and their subsequent recombination are facilitated by the lowering of CBM at the interface to bypass trap states. As quantum information technologies require bright, on-demand, and location-determined single photon emitters, this work paves the way for activating and applying the emitters in a more controllable fashion.

**Acknowledgments**

This work was supported primarily by the U.S. Department of Energy, Office of Science, Basic Energy Sciences, Materials Sciences and Engineering Division under contract no. DE-AC02-05-CH11231, within the sp2-Bonded Materials Program (KC2207), which provided for the development of the project concept and theoretical calculations. CL and HRTEM measurements were provided by The Molecular Foundry, supported under the U.S. Department of Energy, Office of Science, Basic Energy Sciences, Materials Sciences and Engineering Division under contract no. DE-AC02-05-CH11231. Additional support was provided by the U.S. Department of Energy, Office of Science, Basic Energy Sciences, Materials Sciences and Engineering Division under contract no. DE-AC02-05-CH11231, within the van der Waals Bonded Materials Program (KCWF16), which provided for sample fabrication, and within the Theory of Materials Program, which provided theoretical methods and analyses. Preparation of the TEM grid suspended samples was provided by the National Science Foundation, under grant DMR-1807322. C.S. gratefully acknowledges financial support of a Kavli Energy NanoScience Institute Heising-Simons Postdoctoral Fellowship. K.W. and T.T. acknowledge support from the Elemental Strategy Initiative conducted by the MEXT, Japan (Grant Number JPMXP0112101001) and JSPS KAKENHI (Grant Numbers 19H05790 and JP20H00354). W.Z. and F.Z acknowledge the support from the Guangdong Innovation Research Team Project (No. 2017ZT07C062), Guangdong Provincial Key-Lab program (No. 2019B030301001), Shenzhen Municipal Key-Lab program (ZDSYS20190902092905285), and Center for Computational Science and Engineering at Southern University of Science and Technology. J.-H.P. and J.K. acknowledge the support from the US Army Research Office (ARO) MURI under grant no. W911NF-18-1-0431 and the US Army Research Office through the Institute for Soldier Nanotechnologies at MIT, under cooperative agreement no. W911NF-18-2-0048.

**Author contributions**

C.S., A.Z., and S.A. conceived the experiments. A.Z. supervised the whole project. S.G.L. supervised the theoretical studies. C.S. and S.A. designed and performed the regular, *in-situ*, and cryo-cathodoluminescence and HRTEM experiments. F.Z. performed the calculations and, with J.-W. J. and S.G.L., performed the theory analyses. S.K. prepared the samples and fabricated the devices. C.D., A.U. aided in sample preparations. J.P. and J.K. prepared monolayer hBN samples. K.W. and T.T. provided the bulk hBN samples. C.S. composed the paper. All the authors were involved in the paper revisions.

**Competing interests**

Authors declare that they have no competing interests.

**Data and materials availability**

All data are available in the main text or the supplementary materials.

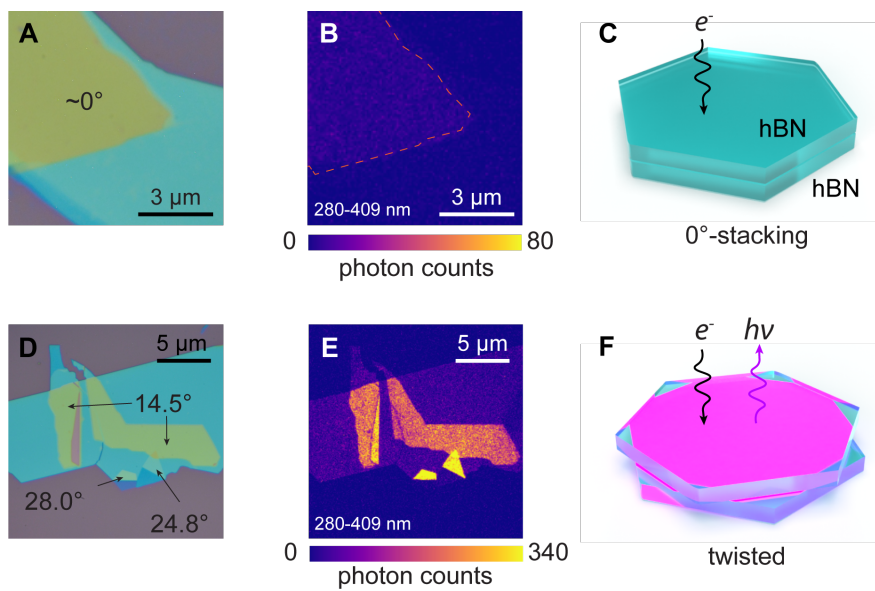

**Fig. 1. Defect emission from T-hBN interface.** The optical images, energy-filtered panchromatic CL intensity map (wavelength window: 280–409 nm), and schematic illustrations of the CL signal in the UV range for stacked hBN flakes with **(A-C)** 0° and **(D-F)** large-angle (> 10°) twist, respectively. The twist angles of T-hBN samples are marked in (D). Both (B) and (E) are acquired at the same conditions.

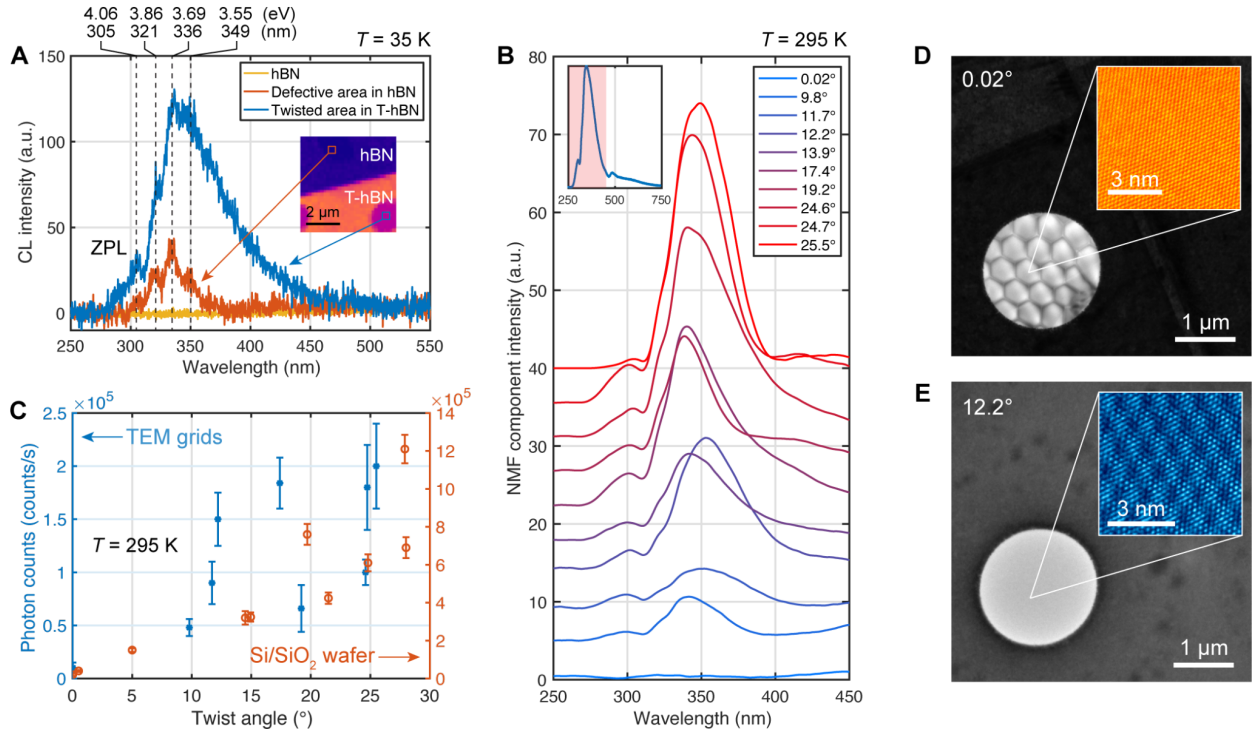

**Fig. 2. Angle-dependent brightness and crystal structures of T-hBN.** **(A)** Cryo-CL spectra of the bulk hBN (yellow), defective area in bulk hBN (red), and the twisted area (blue) on the same sample measured at 35 K. All spectra are from freestanding regions with no substrate underneath. ZPL: zero-phonon line. Inset: a UV intensity mapping shows the locations where the spectra are taken. The round dark holes come from the structure of the TEM grid where hBN/T-hBN is freestanding. **(B)** The NMF decomposition component associated with the CL emission from T-hBN area at room temperature with various twist angles. Inset: a typical full spectrum of the T-hBN ranging from 250 to 750 nm. **(C)** The increasing trend of the integrated UV intensity in 280-409 nm from PMT when the twisting angle approaches 30°. The left and right vertical axes mark the intensity of the freestanding samples on TEM grids (data represented by blue dots) and the samples on $SiO_2$/Si substrate (data represented by orange circles), respectively. All data points represent the photon counts per second per pixel, and error bars represent the standard deviations. The HRTEM images of the lattice structure of freestanding T-hBN show moiré superlattices with **(D)** 0.02°-twist and **(E)** 12.2°-twist, respectively.

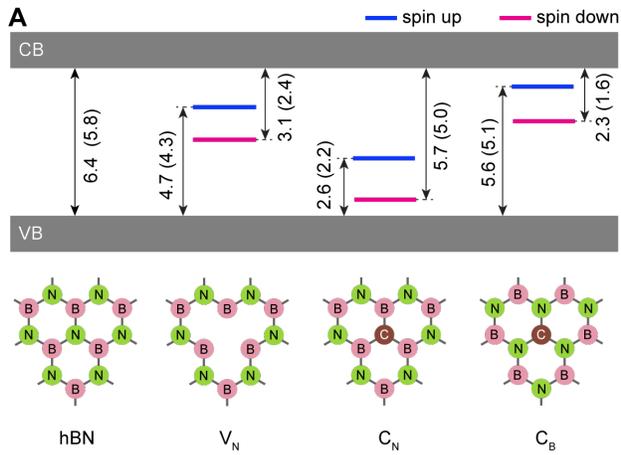
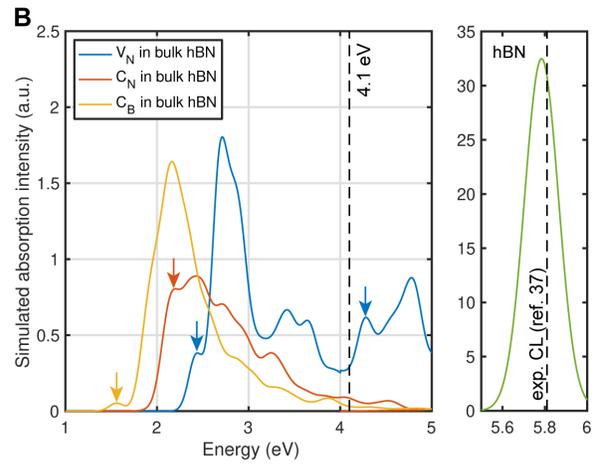
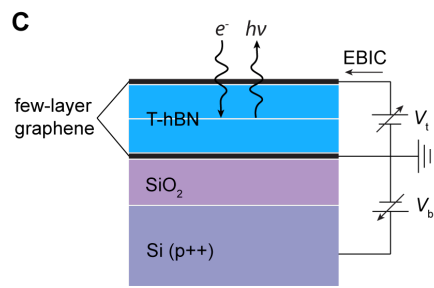
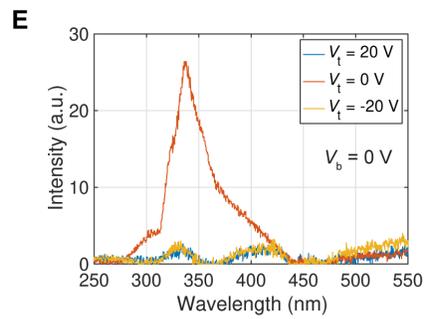
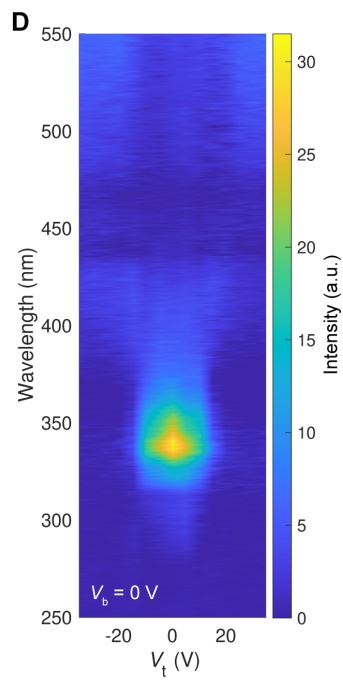
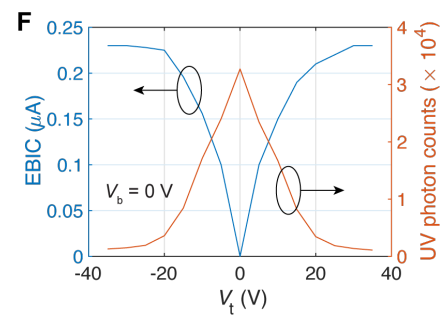
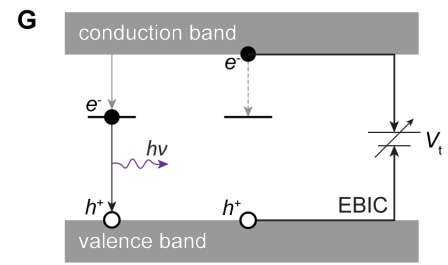
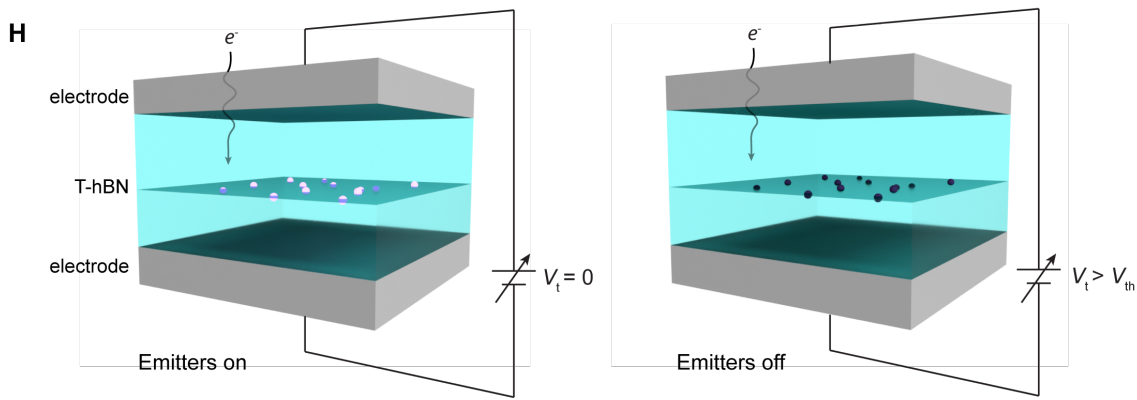

**Fig. 3. Defect type identification and the optical modulation of the emitter *via* external voltage.** **(A)** The energy levels of $V_N$, $C_N$, $C_B$ with respect to CB and VB edges are calculated by *GW* method. Blue and magenta lines mark spin-up and spin-down. The numbers next to the double arrows indicate the energetic distance between the energy levels, and the numbers inside the parentheses indicate the energies of the ground-state excitons after subtracting the electron-hole binding energy estimated by BSE. All the numbers are in units of eV. **(B)** The calculated absorption spectra of $V_N$, $C_N$, and $C_B$ in hBN (left panel), and bulk hBN (right panel) from *ab initio GW*-BSE. The arrows in the left panel mark the lowest energy optically bright excitons, which would typically dominate the spectra during *emission* after excitation. The dashed line in the left panel marks the 4.1-eV ZPL of the color center, and the one in the right panel marks the edge emission of hBN from CL experiment (*37*). **(C)** The schematic device structure of T-hBN encapsulated by two few-layer graphene electrodes. The device is biased by two voltage suppliers with the top voltage, $V_t$, changing the vertical electric field of T-hBN, and the bottom voltage, $V_b$, changing the back gate. EBIC: electron-beam-induced current. **(D)** The color map of the decomposed CL spectra at T-hBN with $V_t$ ranging from -35 V to 35 V, and $V_b = 0$ V, where the spectra of $V_t = 20, 0,$ and -20 V are plotted in **(E)**. **(F)** The anti-correlative EBIC and UV photon intensity as functions of $V_t$, with $V_b = 0$ V. **(G)** The competing behavior of EBIC and UV photon emission for mobile carriers. $V_t$ extracts the mobile carriers away from T-hBN which decreases the formation and hence the recombination rate of excitons (derived from pairing of electrons in localized defect states and holes in valence band states) at the interface. **(H)** A schematic of color center control through external voltages. When no voltage is applied (left), color centers at the twisted interface is bright under the excitation source; when voltage is beyond a threshold voltage, $V_{th}$, the emitters become dark even under excitation (right).

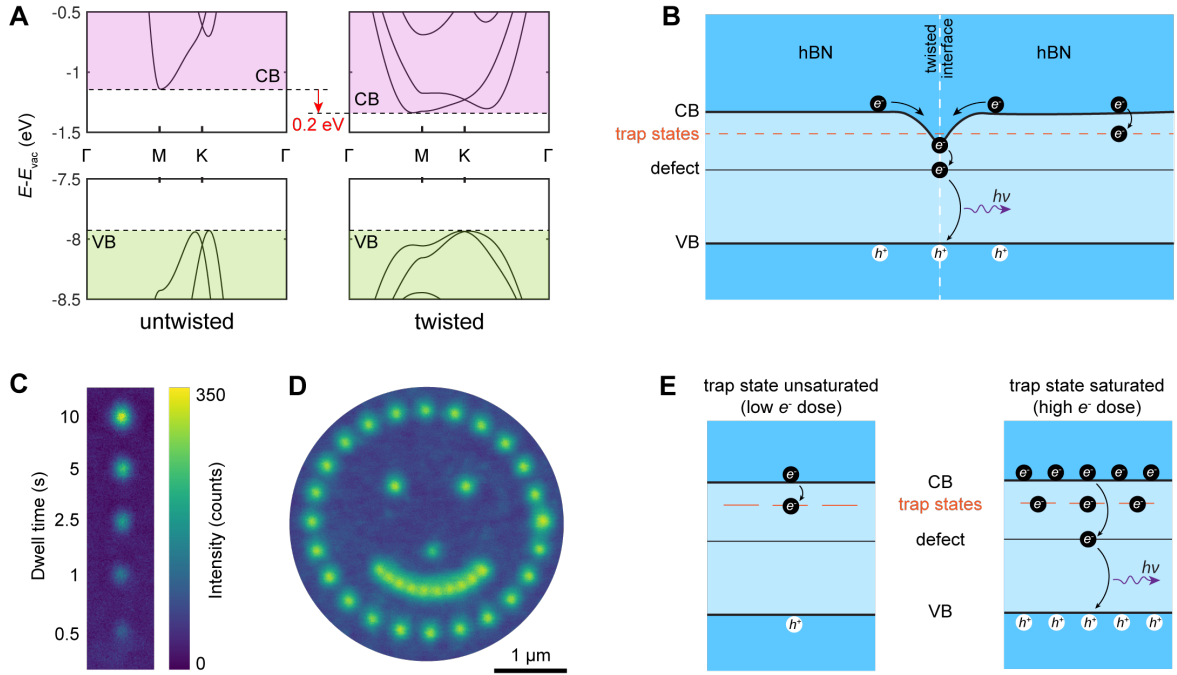

**Fig. 4. Trap-state saturation from moiré potential. (A)** The band structure changes when a bilayer hBN is twisted by 21.78°. The conduction band minimum (CBM) lowers down by 0.2 eV, while the valence band maximum (VBM) is unchanged. **(B)** Upon twisting, the lowering of CBM creates a quantum well for carrier electrons around the interface, moving the local CBM at the interface close to or even below the shallow trap states. The electrons in CB are thus free from trapping and can directly go to the defect level, forming defect excitons with valence band holes. **(C)** An array of color center ensembles was seeded with logarithmically spaced dwell times between 0.5 and 10 seconds. The brightness of the color center ensemble increases as the dwell time increases, which is a feature of saturating trap states. **(D)** The electron beam is placed at a series of locations and dwells for 1 second, resulting in the smiley pattern. **(E)** A schematic of the emission enhancement coming from "trap-state saturation". Left panel: electrons are trapped by trap states without a radiative decay. Right panel: trap-state saturation leads to radiative decays with UV emission.